\documentclass[aps,prl,reprint,groupedaddress,preprintnumbers]{revtex4-2}
\usepackage{amsmath,bbm,graphicx,diagbox}

\newcommand{\la}{\leftrightarrow}

\DeclareMathSymbol{\mg}{\mathrel}{symbols}{"1D}

\newcommand{\bes}{\begin{split}}
\newcommand{\ees}{\end{split}}

\newcommand{\gb}{\beta}

\newcommand{\gm}{\mu}

\newcommand{\gl}{\lambda}

\newcommand{\gps}{\psi}

\newcommand{\gch}{\chi}
\newcommand{\ra}{\rightarrow}

\newcommand{\der}{\partial}

\newcommand{\beq}{\begin{equation}}
\newcommand{\eeq}{\end{equation}}
\newcommand{\barr}{\begin{array}}
\newcommand{\earr}{\end{array}}
\newcommand{\equ}[1]{\begin{gather} #1 \end{gather}}

\newcommand{\arry}[2]{\begin{array}{#1} #2 \end{array}}

\newcommand{\sfrac}[2]{\mbox{$\frac{#1}{#2}$}}
\newcounter{oldcounter}

\newcommand{\bder}{\bar\partial}
\newcommand{\bff}{{\bar f}}

\newcommand{\bgb}{{\bar\beta}}

\newcommand{\bgl}{{\bar\lambda}}

\newcommand{\Bgb}{{\boldsymbol \beta}}

\newcommand{\Bgx}{{\boldsymbol \xi}}

\newcommand{\Intr}{\mathbbm{Z}}

\newcommand{\ba}[2]{\[\begin{array}{#2}\label{#1}}
\newcommand{\ea}{\end{array}\]}
\newcommand{\be}{\begin{equation}}
\newcommand{\ee}{\end{equation}}
\newcommand{\bea}{\begin{eqnarray}}
\newcommand{\eea}{\end{eqnarray}}

\newcommand{\sm}{{\,\mbox{-}}}

\newcommand{\ztwo}{\Intr_2\!\times\!\Intr_2}

\newcommand{\FermSymBosCoordinates}{ 
\begin{table}[h!]
\caption{\label{tb:FermSymBosCoordinates}
Fermionic symmetry induced bosonic coordinate field transformations. (Only non--inert fields are given.) 
}
\begin{ruledtabular}
\begin{tabular}{cc}
\textrm{Fermionic symmetry} & \textrm{Action on left--moving bosons}
\\ \hline 
$(2a\!\sm\! 1\, 2b\!\sm\!1) (2a\, 2b)$
&
$X_\text{L}^{a} \la   X_\text{L}^{b}$
\\[0.5ex] 
$(2a)$ & 
$X_\text{L}^a \ra - X_\text{L}^a$ 
\\ \hline 
$(2a\!\sm\! 1)$ & 
$X_\text{L}^a \ra  \pi - X_\text{L}^a$ 
\\[0.5ex] 
$(2a\!\sm\! 1\, 2a)$
& 
$X_\text{L}^a \ra   \sfrac12\pi - X_\text{L}^a$ 
\end{tabular}
\end{ruledtabular}
\end{table}
}

\newcommand{\FermSymBosBoundaryConditions}{
\begin{table}[h!]
\caption{\label{tb:FermSymBosBoundaryConditions}
Fermionic symmetry induced bosonic boundary condition mappings. (Only non--inert entries of the vectors $v_\text{L}$ and $V_\text{L}$ modulo integral vectors are given.) 
}
\begin{ruledtabular}
\begin{tabular}{cc}
\textrm{Fermionic symmetry} & \textrm{Action on twist and shift entries}
\\ \hline 
$(2a\!\sm\! 1\,2b\!\sm\! 1)(2a\, 2b)$
&
$v_\text{L}^{a} \la   v_\text{L}^{b}\,,~V_\text{L}^{a} \la  V_\text{L}^{b}$
\\[0.75ex] 
$(2a)$ & 
$v_\text{L}^a \ra - v_\text{L}^a +2\, V_\text{L}^a\,,~ V_\text{L}^a \ra V_\text{L}^a$
\\ \hline 
$(2a\!\sm\! 1\, 2a)$
& 
$v_\text{L}^a \ra - v_\text{L}^a\,,~ V_\text{L}^a \ra V_\text{L}^a- v_\text{L}^a$ 
\\[0.75ex] 
(2a\, 2b) & 
$\arry{l}{v_\text{L}^a \ra v^b_\text{L} + V^a_\text{L} - V^b_\text{L}\,,~ V^a_\text{L} \ra V^a_\text{L}\,,  
\\[-0.5ex]  
v_\text{L}^b \ra v^a_\text{L}+V^b_\text{L}-V^a_\text{L}\,,~ V^b_\text{L} \ra V^b_\text{L}}$ 
\\[3ex] 
(2a\!\sm\! 1\, 2b\!\sm\!1) & 
$\arry{l}{v_\text{L}^a \ra v^a_\text{L} + V^a_\text{L} - V^b_\text{L}\,,~ V^a_\text{L} \ra V^b_\text{L}\,,  
\\[-0.5ex]  
v_\text{L}^b \ra v^b_\text{L}+V^b_\text{L}-V^a_\text{L}\,,~ V^b_\text{L} \ra V^a_\text{L}}$ 
\\[3ex] 
$(2a\!\sm\! 1)$ & 
$v_\text{L}^a \ra v_\text{L}^a - 2\, V_\text{L}^a\,,~ V_\text{L}^a \ra -V_\text{L}^a$ 
\end{tabular}
\end{ruledtabular}
\end{table}
}

\newcommand{\VectorsExample}{ 
\begin{table}[t!]
\caption{\label{tb:VectorsExample}
Fermionic basis vectors $\Bgb$ with the twists $v$ and shifts $V$ corresponding to $\mathbf{1}\!\!-\!\!\Bgb$ obtained via~\eqref{eq:HolomorphicTranslationFermionicBosonicBoundaryConditions}
and~\eqref{eq:AntiHolomorphicTranslationFermionicBosonicBoundaryConditions}. 
}
\begin{ruledtabular}
\begin{tabular}{l|ll}
\textrm{Fermionic basis vectors}  & 
\textrm{Twist  and} & 
\textrm{shift vectors}
\\ 
$\Bgb$ & 
$v\!\!=\!\!(v_\text{R}|v_\text{L})$ & 
$V\!\!=\!\!(V_\text{R}|V_\text{L})$ 
\\ \hline 
$\mathbf{1}\!\!=\!\!\{\psi^{1\ldots 4} \chi^{1\dots 4} y^{1\dots 4} w^{1\dots 4} | \bar{f}^{1\ldots 40}\}$ & $(0^4| 0^{20})$ & $\sfrac 12(1^4| 1^{20})$
\\
$\mathbf{S}\!\!=\!\!\{\psi^{1\ldots 4}\chi^{1\dots 4}\}$ & $(0^4| 0^{20})$ & $(0^4| 0^{20})$ 
\\ 
$\Bgx\!\!=\!\!\{\bar{f}^{9 \dots 40}\}$ & $(0^4| 0^{20})$ &  $\sfrac 12(0^4| 0^4 1^{16})$
\\ 
$\mathbf{b}\!\!=\!\!\{\chi^{1\ldots 4}w^{1,\ldots 4} | \bar{f}^{1\ldots 4} \bar{f}^{9\ldots 12}\}$ & $\sfrac 12(1^4| 0^{20})$ & $\sfrac 12(0^4| 1^2 0^2 1^2 0^{14})$
\end{tabular}
\end{ruledtabular}
\end{table}
}

\newcommand{\SimpleWebExample}{
\begin{figure}[h!]
\[
  \includegraphics[width=0.95\linewidth]{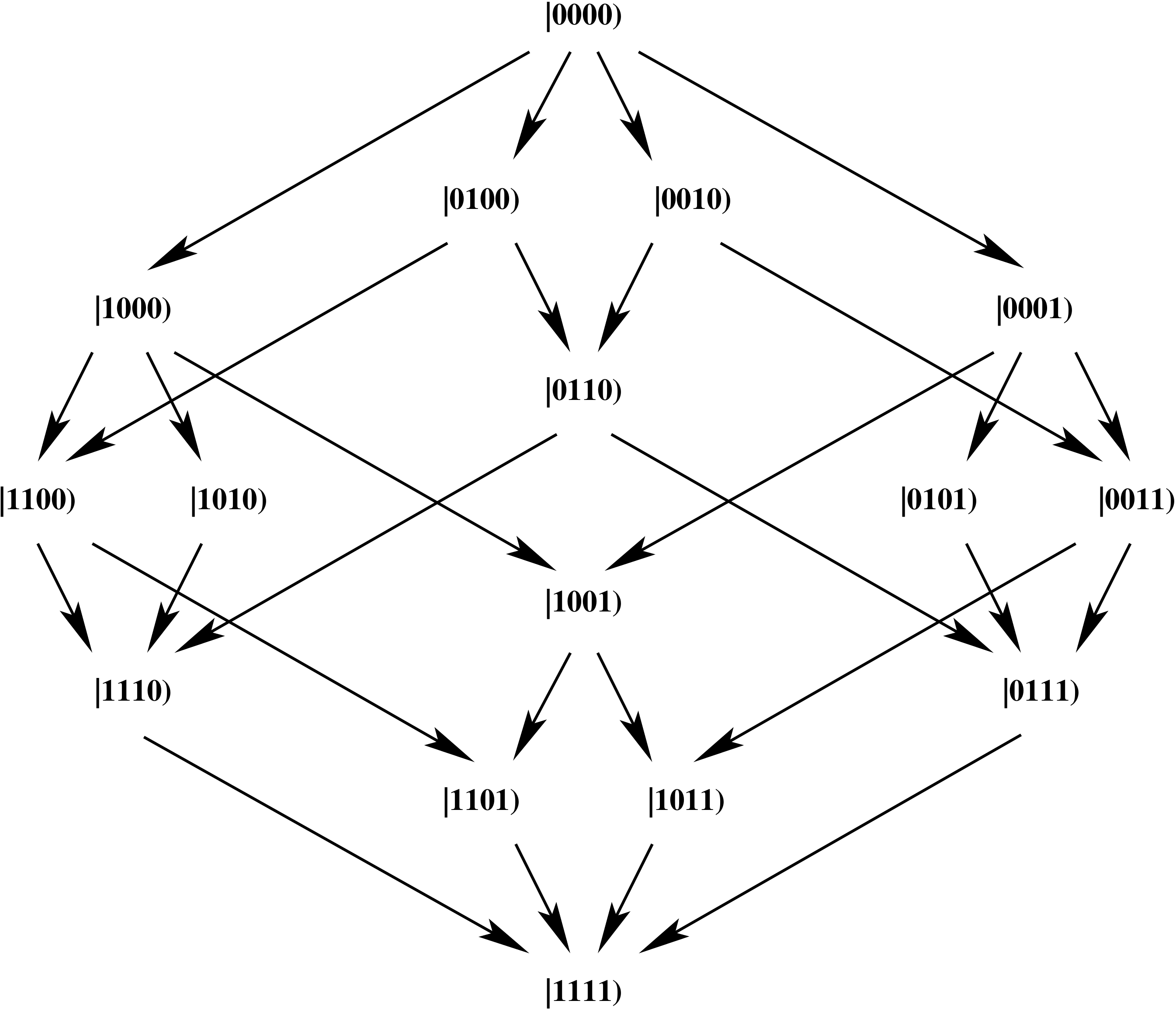}
\]
\caption{\label{fg:SimpleWebExample} 
Web of T--folds associated with the fermionic model given in Table~\ref{tb:VectorsExample} in which only $\mathbf{b}$ acts as a twist~\eqref{eq:PermutatedTwists}.  
}
\end{figure}
}

\newcommand{\FrequencyTableExample}{
\begin{table}[h!]
\caption{\label{tb:FrequencyTableExample} 
The number of $\mathbf{b}$ and $\Bgx$ twists for the inequivalent T--folds associated with the fermionic model in Table~\ref{tb:VectorsExample}. 
}
\begin{ruledtabular}
\begin{tabular}{ c | ccccc c }
\backslashbox{$\boldsymbol{~\mathbf{b}~}$}{$\boldsymbol{\Bgx}$} 
 & \textbf{0} & \textbf{2} & \textbf{4} & \textbf{6} & \textbf{8} & \textbf{Sum} \\
 \cline{1-7} 
\textbf{0} & 1 & 2 & 3 & 2 & 1 & \textbf{9} \\
\textbf{2} & 2 & 11 & 18 & 12 & 3 & \textbf{46} \\
\textbf{4} & 3 & 18 & 32 & 19 & 6 & \textbf{78} \\
\textbf{6} & 2 & 12 & 19 & 18 & 7 & \textbf{58} \\
\textbf{8} & 1 & 3 & 6 & 7 & 5 & \textbf{22} \\ 
\textbf{Sum} & \textbf{9} & \textbf{46} & \textbf{78} & \textbf{58} & \textbf{22} & \textbf{213} \\ 
\end{tabular}
\end{ruledtabular}
\end{table} 
}

\begin{document}

\preprint{LTH-1343}
\preprint{CERN-TH-2023-115}

\title{Free fermionic webs of heterotic T--folds}


\author{Alon E. Faraggi}
\email[]{alon.faraggi@liverpool.ac.uk}
%\thanks{}
\affiliation{Dept.\ of Mathematical Sciences, University of Liverpool, Liverpool
L69 7ZL, UK\\
and CERN, Theoretical Physics Department, CH–-1211 Geneva 23, Switzerland}

\author{Stefan Groot Nibbelink}
\email[]{s.groot.nibbelink@hr.nl}
%\thanks{}
\affiliation{School of Engineering and Applied Sciences and 
Research Centre Innovations in Care, 
Rotterdam University of Applied Sciences,
Postbus 25035, 3001 HA Rotterdam, the Netherlands}

\author{Benjamin Percival}
\email[]{b.percival@liverpool.ac.uk}
%\thanks{}
\affiliation{Dept.\ of Mathematical Sciences, University of Liverpool, Liverpool
L69 7ZL, UK}

%\affiliation{}

\date{\today}

\begin{abstract}
Moduli stabilisation is key to obtaining phenomenologically viable string models. Non--geometric compactifications, like T--duality orbifolds (T--folds), are capable of freezing many moduli. However, in this Letter we emphasise that T--folds, admitting free fermionic descriptions, can be associated with a large number of different T--folds with varying number of moduli, since the fermion pairings for bosonisation are far from unique. Consequently, in one description a fermionic construction might appear to be asymmetric, and hence non--geometric, while in another it admits a symmetric orbifold description. We introduce the notion of intrinsically asymmetric T--folds for fermionic constructions that do not admit any symmetric orbifold description after bosonisation. Finally, we argue that fermion symmetries induce mappings in the bosonised description that extend the T--duality group.
\end{abstract}

%\keywords{LTH-XXXX}

\maketitle

\section{\label{sc:Introction} Introduction}

String theory realises a unification of gravity,  gauge interactions and their charged matter via the properties of Conformal Field Theories (CFTs) residing on its two dimensional (2D) worldsheet. 
Heterotic strings on toroidal orbifolds~\cite{Dixon:1985jw,Dixon:1986jc} led to some of the most realistic string--derived models to date~\cite{Faraggi:1989ka,Lebedev:2006kn,Blaszczyk:2009in}.
However, orbifolds and other geometrical backgrounds result in free moduli (such as the metric, B--field or Wilson lines) on which detailed physics, like gauge and Yukawa couplings, depend. 

Strings on tori and their orbifolds admit exact quantisation. This was instrumental in the discovery of T--dualities~\cite{Duff:1989tf}, like the famous $R \ra 1/R$ duality, which sets the effective minimum of the radius $R$ equal to the string scale. 
Investigations of string backgrounds had a profound impact on mathematics as mirror symmetry showed, which was argued to be a form of T--duality~\cite{Strominger:1996it}. 

Modding out T--duality symmetries may lead to exotic non--geometric backgrounds~\cite{Dabholkar:2005ve,Grana:2008yw}, dubbed T--folds. 
Hence, the landscape of string vacua may be much vaster than suggested by geometrical compactifications alone. 
Even though non--geometric constructions have been studied far less than their geometric counterparts, they may be vital for phenomenological string explorations, as they are capable of freezing many moduli. 

Such T--folds may have different actions on their left-- and right--moving bosonic coordinate fields, and are thus referred to as asymmetric orbifolds~\cite{narain_87,ibanez_88}.
If only order two symmetries are modded out, an alternative fermionic description may be obtained by bosonisation, a CFT equivalence of chiral bosons and fermions in 2D~\cite{DiFrancesco:1997nk}. 
This led to a detailed dictionary between these two formulations explicated for symmetric $\ztwo$ orbifolds~\cite{Athanasopoulos:2016aws}.
Asymmetric boundary conditions in the fermionic formalism have profound phenomenological consequences,  such as doublet--triplet splitting mechanism~\cite{Faraggi:1994cv,dtsm}, Yukawa coupling selection~\cite{yukawa}, and moduli fixing~\cite{moduli}.

Although a similar dictionary for asymmetric orbifolds is not this letter's aim, heterotic bosonisation ambiguities suggest identifications of seemingly unrelated T--folds. 
This sheds new light on non--geometric moduli stabilisation. 
Fermionic symmetries parameterising these ambiguities, suggest an extension of the T--duality group.

\section{\label{sc:Tfolds} Order two bosonic T--fold models}

The bosonic formulation of the heterotic string~\cite{Gross:1985fr} describes $d$--dimensional Minkowski space by coordinate fields $x=(x^{\gm=2\ldots d-1})$ in the light--cone gauge. 
The internal coordinate fields $X=(X_\text{R} | X_\text{L})$,  with right-- and left--chiral parts,  $X_\text{R}=(X_\text{R}^{i=1\ldots D})$ and $X_\text{L} = (X_\text{L}^{a=1\ldots D+16})$, $D=10-d$, are subject to torus periodicities 
\equ{ \label{eq:TorusPeriodicities}
X \sim X + 2\pi\, N\,, 
\qquad 
N \in \Intr^{2D+16}\,. 
}
The worldsheet supersymmetry current,
\equ{\label{eq:SupercurrentBosonicForm}
T_\text{F}(z)=i\,\psi_\mu \partial x^\mu +i\, \chi^i \partial X_\text{R}^i\,,
}
involves the real holomorphic superpartners $\gps =(\gps^\mu)$ and $\gch = (\gch^i)$ of $x$ and $X_\text{R}$, respectively. 
Here, $(\bder)\der$ denotes (anti--)holomorphic worldsheet derivative and repeated indices are summed over. 

An order two generator, defining the orbifold action
\equ{ \label{eq:OrbiActions}
X \sim e^{2\pi i\, v}\, X - 2\pi\, V\,, 
}
with 
\(
v = (v_\text{R} | v_\text{L}),\, 
V = (V_\text{R} | V_\text{L})
\in 
\frac 12\, \Intr^{2D+16}
\), 
is called a shift, a twist, or a roto--translation, if $V\not\equiv 0 \equiv v$, $v\not\equiv 0 \equiv V$, or $v, V\not\equiv 0$, respectively. ($\equiv$ means equal up to integral vectors.) 

An orbifold is called \textit{symmetric} if there is a basis such that the left-- and right--twist parts are equivalent according to
\equ{
v_\text{L} \equiv (v_\text{R}, 0^{16})\,, 
}
for all its generators simultaneously~\cite{Ibanez:1987pj,GrootNibbelink:2017usl,GrootNibbelink:2020dib}, and \textit{asymmetric} if no such basis exists. 
The addition of $0^{16}$ is essential, as the vectors $v_\text{L}$ and $v_\text{R}$ have unequal lengths.

\section{\label{sc:FermionicModels} Real free fermionic models}

In the free fermionic formulation~\cite{Antoniadis1987a,Kawai1987,AB} the internal degrees of freedom are described by real holomorphic fermions $f=(y,w)$ with $y=(y^i)$ and $w=(w^i)$ and real anti--holomorphic fermions $\bff = (\bff^{u=1\ldots 2D+32})$. 
Worldsheet supersymmetry is realised non--linearly 
\equ{\label{eq:Supercurrent}
T_\text{F}(z)=i\psi_\mu \der x^\mu +i \chi^i y^i w^i\,.
}
A fermionic model is defined by a set of basis vectors with entries $0$ or $1$ for real fermions. Each basis vector $\Bgb = (\gb |\bgb)$ defines boundary conditions
\equ{\label{eq:FermionicBoundaryConditions}
  f \sim -e^{\pi i\, \gb}\, f~, 
  \qquad 
  \bff \sim -e^{\pi i\, \bgb}\, \bff\,.  
}

\section{\label{sc:Bosonisations} Bosonisations} 

\subsection{Holomorphic bosonisation}

Assuming that the fermions $\gch$ are identical in the supercurrents~\eqref{eq:SupercurrentBosonicForm} and~\eqref{eq:Supercurrent} and they generate the same worldsheet supersymmetry in the bosonic and fermionic descriptions, bosonisation uniquely relates the currents 
\begin{subequations} \label{eq:HolomorphicBosonisation} 
\equ{ \label{eq:HolomorphicCurrentBosonisation}
J^i =\ :\! (\gl^{i})^* \gl^i\!:\ =\ :\! y^i w^i \!:\ \cong i\, \der X_\text{R}^i\, ,
}
and complex fermions 
\equ{ \label{eq:HolomorphicFermionBosonisation}
\gl^i = \frac 1{\sqrt 2}( y^i + i\, w^i) \cong \ :\! e^{i\,X_\text{R}^i} \!:
}
\end{subequations} 
to normal ordered exponentials of chiral bosons. Here $\cong$ emphasises that these expressions are not identities but rather that both sides have identical operator product expansions in either formulation.

The bosonisation formulae relates the boundary conditions in both descriptions. The torus periodicities~\eqref{eq:TorusPeriodicities} reflect the $2\pi$ ambiguities of $X_\text{R}$ in the complex exponentials~\eqref{eq:HolomorphicFermionBosonisation}. Comparing the orbifold conditions~\eqref{eq:OrbiActions} of the right--moving bosons $X_\text{R}$ with boundary conditions~\eqref{eq:FermionicBoundaryConditions}
of the holomorphic fermions $y$ and $w$ in~\eqref{eq:HolomorphicBosonisation} leads to the following identifications: 
\equ{ \label{eq:HolomorphicTranslationFermionicBosonicBoundaryConditions}
v_\text{R} = \sfrac12 \gb(w) - \sfrac12 \gb(y)\,, 
\qquad 
V_\text{R} = \sfrac12 (1^D) - \sfrac 12 \gb(y)\,. 
}

\subsection{Anti--holomorphic bosonisation}

Contrary to the holomorphic side, the pairing of the anti--holomorphic fermions is arbitrary. Associating odd and even fermion labels to the real and imaginary parts of complex fermions results in an anti--holomorphic bosonisation procedure given by: 
\begin{subequations}  \label{eq:AntiHolomorphicBosonisation}
\equ{ \label{eq:AntiHolomorphicCurrentBosonisation}
\overline{J}^{a} =\ :\! (\bgl^{a})^* \bgl^{a} \!:\ =\ :\! \bar{f}^{2a-1} \bar{f}^{2a} \!:\ \cong i\, \bder X_\text{L}^{a}~,
}
with  
\equ{ \label{eq:AntiHolomorphicFermionBosonisation} 
\bgl^{a} = \frac 1{\sqrt 2} (\bff^{2a-1} + i \bff^{2a} ) \cong\ :\! e^{i X_\text{L}^{a}} \!:~,
}
\end{subequations} 
for $a=1,\ldots D+16$.

Then by similar arguments as above, the torus periodicities~\eqref{eq:TorusPeriodicities} for $X_\text{L}$ follow. And splitting $\bgb = (\bgb_\text{o}, \bgb_\text{e})$ in two $(D+16)$--dimensional vectors, $\bgb_\text{o}=(\bgb^{1,3\ldots 2D+31})$ and $\bgb_\text{e}=(\bgb^{2,4\ldots 2D+32})$ leads to the identifications: 
\equ{ \label{eq:AntiHolomorphicTranslationFermionicBosonicBoundaryConditions}
v_\text{L} = \sfrac12 \bgb_\text{e} - \sfrac12 \bgb_\text{o}\,,
\qquad
V_\text{L} = \sfrac12 (1^{D+16}) - \sfrac 12 \bgb_\text{o}\,. 
}

\FermSymBosCoordinates
\FermSymBosBoundaryConditions

\section{\label{sc:TdualityExtension} Extension of the T--duality group}

\subsection{Fermionic inversions and permutations}

On the anti--holomorphic side, the fermionic symmetries contain inversions and permutations: $(u)$ denotes the fermion inversion $\bff^u \ra -\bff^u$. The permutation $(u_1 \ \cdots \ u_p)$ acts as $\bff^{u_1} \ra \bff^{u_2} \cdots \ra \bff^{u_p} \ra \bff^{u_1}$ leaving the remaining fermions inert. 
The permutation group contains elements which consist of multiple factors like this provided their entries are all distinct. 
It is generated by permutations of two elements $(uv)$. 
The induced fermionic symmetry actions within the bosonic formulation can be identified using the bosonisation~\eqref{eq:AntiHolomorphicBosonisation}.

\subsection{Induced bosonic coordinate transformations}

The fermionic symmetries, that leave these fermion bosonisation pairs intact, realise mappings of the bosonic coordinate fields $X_\text{L}$ to themselves. Their generators and their realisations on the bosonic coordinates are listed in Table~\ref{tb:FermSymBosCoordinates}. The bosonic transformations above the middle line of this table are part of the T--duality group, while those below involve translations as well. 

\subsection{Induced mappings of bosonic boundary conditions}

Other fermionic symmetries break up fermion bosonisation pairs and hence correspond to mappings between different coordinate fields between which no obvious coordinate transformation exists. 
However, all fermionic symmetries, generated by inversions and permutations, map the boundary conditions of one or\-bifold theory to another. 
The mappings induced by the generators of the fermionic symmetries are collected in Table~\ref{tb:FermSymBosBoundaryConditions}. 
The transformations induced by permutations $(2a\,2b)$ and $(2a\!\sm\! 1\, 2b\!\sm\! 1)$ combined (in whatever order) leads to the boundary condition mapping associated with  $(2a\!\sm\! 1\, 2b\!\sm\! 1)(2a\,2b)$ as the group property would suggest. 
Since some actions can be interpreted as T--duality transformations, while others cannot, this hints at an extension of the T--duality group.  

The Table~\ref{tb:FermSymBosBoundaryConditions} mappings $(2a\!\sm\!1\, 2a)$, $(2a\!\sm\!1\, 2b\!\sm\!1)$ and $(2a\, 2b)$ are of special significance: they mix the twist and shift vector entries. The action of $(2a\!\sm\!1\, 2a)$ recalls that the shift part of a roto--translation in directions, where the twist acts non--trivially, can be removed via the associated coordinate transformation in Table~\ref{tb:FermSymBosCoordinates}. The actions $(2a\!\sm\!1\, 2b\!\sm\!1)$ and $(2a\, 2b)$ imply that a pure shift boundary condition can be turned into a roto--translation. By combining these mappings, a web of equivalent (mostly asymmetric) orbifold theories emerge. 

Since all these T--folds are just different bosonic representations of the same fermionic theory, their physical properties are identical, even though they may not look alike. For example, their modular invariance conditions may seem to disagree, as the number of non--zero entries in the twist vectors under mappings, like $(2a\, 2b)$, change. However, since only the part of the shift  of~\eqref{eq:OrbiActions}, on which the twist acts trivially, takes part in the modular invariance condition~\cite{GrootNibbelink:2020dib}, their consistency conditions are numerically identical.

\subsection{A free fermionic T--fold web}

\VectorsExample

The basis vectors, $\Bgb$, 
for a simple illustrative 6D fermionic model are given in Table~\ref{tb:VectorsExample} together with associated twist and shift vectors using the odd--even pairings~\eqref{eq:AntiHolomorphicBosonisation}. Within this bosonisation the model is understood as an asymmetric orbifold. The interpretation may change by applying fermionic symmetries.

The permutations  
$(2\, 6)^{p_1} (4\, 8)^{p_2} (10\,14)^{p_1} (12\,16)^{p_4}$ with $p_i=0,1$, map the twist vector $v_\text{L}(\mathbf{1}\!-\!\mathbf{b}) = \sfrac12 (0^{20}) \ra$ 
\equ{ \label{eq:PermutatedTwists} 
|p_1p_2\, p_3p_4) = \sfrac 12 (p_1 p_2 p_1 p_2\, p_3 p_4 p_3 p_4\, 0^{12})\,,
}
while the other twists and shifts remain the same, since $(2a\, 2b)$ leave $V_\text{L}$ inert (see Table~\ref{tb:FermSymBosBoundaryConditions}). 
When these permutations are successively switched on, the T--fold web, given in Figure~\ref{fg:SimpleWebExample}, is obtained. 

For the cases with two non--zero $p_i$, \eqref{eq:PermutatedTwists} implies that $v_\text{L} = (v_\text{R},0^{16})$ possibly up to a change of basis. Thus the resulting bosonic models are interpreted as symmetric orbifolds. 
In particular, the model obtained after the fermionic permutation $(2\, 6) (4\, 8)$ is conventionally considered as the bosonic representation of this fermionic model in which  $\Bgx$ just separates out the $SO(32)$ gauge group, while for all the other cases with two non--zero $p_i$, $\Bgx$ acts as an asymmetric Wilson line. 

Table~\ref{tb:FrequencyTableExample} provides an overview of all inequivalent T--fold models associated with this fermionic model. 
It indicates in how many directions $\mathbf{b}$ and $\Bgx$ act as left--moving twists. 
Apart from the sixteen models depicted in Figure~\ref{fg:SimpleWebExample} (of which nine are inequivalent), $\Bgx$ has an asymmetric twist action as can inferred from this table. 
The total number of inequivalent T--fold models associated with the fermionic basis vectors given in Table~\ref{tb:VectorsExample} is $213$. 
This number rapidly increases for fermionic models defined with more basis vectors. For example, for the fermionic model in which $\Bgx$ is split into $\Bgx_1$ and $\Bgx_2$ the number of inequivalent bosonisations becomes $11\,273$ and  for the NAHE set~\cite{Ferrara:1987jr,Antoniadis:1987tv,Faraggi1993} is $85\,735$.

\SimpleWebExample

\FrequencyTableExample

\section{\label{sc:Moduli} Moduli} 

The unfixed Narain moduli ($m_{ij}=g_{ij}+b_{ij}$ with metric $g_{ij}$, B--field $b_{ij}$ and Wilson lines $m_{ix=1\ldots 16}$) of a T--fold correspond to the operators,
\equ{ \label{eq:BosonicDeformations} 
m_{ia}\, \der X_\text{R}^i \, \bder X_\text{L}^a\,, 
}
left inert by~\eqref{eq:OrbiActions}. 
Symmetric orbifolds always leave at least the diagonal metric moduli $m_{ii} =g_{ii}$ free, asymmetric orbifolds may fix all moduli.

This would suggest that the number of frozen moduli may vary dramatically depending on which bosonic description of a given fermionic model is used. 
There is no paradox here either: 
the unfixed scalar deformations of the fermionic model can be identified by the Thirring interactions, 
\equ{ \label{eq:FermionicDeformations} 
m_{iuv}\,  y^iw^i\, \bff^u\! \bff^v\,,
}
left inert by~\eqref{eq:FermionicBoundaryConditions}. 
Thus, the total number of massless untwisted scalars is bosonisation independent, and therefore identical in any bosonic realisation. 
Which of them are interpreted as free Narain deformations, however, does depend on the choice of bosonisation, as $X_\text{L}$ in~\eqref{eq:BosonicDeformations} does.

\section{\label{sc:IntrinsicallyAsymTfolds} Intrinsically asymmetric T--folds}

The previous section showed that whether a real fermionic model should be considered as a symmetric or asymmetric model is very much bosonisation dependent. 
A free fermionic model is called \textit{intrinsically asymmetric} if for any bosonisation it corresponds to an asymmetric orbifold. 
An intrinsically asymmetric T--fold is a bosonic model associated with an intrinsically asymmetric  fermionic model.

In light of the observation below~\eqref{eq:BosonicDeformations}, a fermionic model that admits a symmetric interpretation has at least inert Thirring interactions~\eqref{eq:FermionicDeformations} with different $u\neq v$ for each $i$. If  not, the fermionic model is intrinsically asymmetric
and hence in any bosonic realisation all Narain moduli are frozen. 
This is, in particular, the case when no Thirring interactions~\eqref{eq:FermionicDeformations} are invariant under~\eqref{eq:FermionicBoundaryConditions}. 
An example of such a model is given in ref. \cite{slm1}.

Simple examples of intrinsically asymmetric free fermionic models can be obtained by taking basis vectors that act as purely holomorphic twists.
(For example, consider the twist basis vector $\mathbf{b} =  \big\{\chi^{1\ldots 4},y^{1\ldots 4} \big\}$ in 6D or $\mathbf{b}_1 =  \big\{\chi^{1\ldots 4},y^{1\ldots 4} \big\}$ and $\mathbf{b}_2 =  \big\{\chi^{3\ldots 6},y^{3\ldots 6} \big\}$ in 4D.) 
As there are no invariant Thirring interactions~\eqref{eq:FermionicDeformations} possible, the corresponding T--fold models are necessarily intrinsically asymmetric.

\section{\label{sc:Discussion} Discussion} 

This letter focused on heterotic T--folds that admit fermionic descriptions. 
Even though the key observation that bosonisation in a fermionic CFT is not unique is not new, its striking consequences seem not to have been appreciated so far: a single free fermionic model can be associated with a large number of seemingly unrelated bosonic theories. 
Some may admit a symmetric orbifold interpretation while most others are asymmetric, but in many different ways. 

In light of this, studies of non--geometric constructions, and T--folds in particular, may need to be revised, since seemingly different non--geometries may, in fact, be equivalent. 
In particular, in the bosonic orbifold literature it would be inconceivable that symmetric and asymmetric orbifolds can be identified. 
Moreover, the number of frozen moduli turns out to be a bosonisation dependent quantity; only the total number of massless untwisted scalars is identical in any description. 
Only for an intrinsically asymmetric T--fold all Narain moduli are fixed in any bosonic description. 
In addition, the induced bosonic actions of fermionic symmetries hints at an extension of the T--duality group of toroidal and $\Intr_2$ orbifold compactifications. 

The findings presented here were derived at free fermionic points. 
However, the induced transformations of the bosonic boundary conditions by the fermionic symmetries may be considered without referring to the fermionic description. Hence, it is an interesting question whether the suggested extension of the T--duality group discussed above is a general duality symmetry of string theory or exists at free fermionic points only.

Our analysis is partially motivated by quasi-realistic model building using the free fermionic formulation, see e.g.\ \cite{slm1}, to give rise to some central features of the Standard Model and its supersymmetric extensions, such as the existence of three generations charged under the Standard Model gauge group with potentially viable Yukawa couplings to Higgs doublets. 
While this paper focused on the moduli of the internal manifolds, there exist free fermionic models in which the moduli space is further restricted~\cite{cleaverwithmanno}, which shows the need for a deeper understand the moduli space in these quasi--realistic example which our analysis may provide. 
Moreover, the methods adopted in the supersymmetric cases considered here can also be utilised in non--supersymmetric string constructions as well as in tachyon free models that are obtained from compactifications of tachyonic ten dimensional  string vacua~\cite{stable}.
In this respect, the understanding of the correspondence between the fermionic and bosonic representations of string vacua is essential to obtain a more profound understanding of the string dynamics at the Planck scale. 

\section{acknowledgments}
AEF would like to thank the CERN theory division for hospitality and support. 
SGN would like to thank the organizers of the workshop Gauged Linear Sigma Models @ 30 for their kind invitation and the warm hospitality of the Simons Center at Stony Brook University during this event where part of this work was done. 
The work of BP is supported 
by EPSRC grant EP/W522399/1.

\bibliography{papers}

\providecommand{\noopsort}[1]{}\providecommand{\singleletter}[1]{#1}%
\begin{thebibliography}{30}%
\makeatletter
\providecommand \@ifxundefined [1]{%
 \@ifx{#1\undefined}
}%
\providecommand \@ifnum [1]{%
 \ifnum #1\expandafter \@firstoftwo
 \else \expandafter \@secondoftwo
 \fi
}%
\providecommand \@ifx [1]{%
 \ifx #1\expandafter \@firstoftwo
 \else \expandafter \@secondoftwo
 \fi
}%
\providecommand \natexlab [1]{#1}%
\providecommand \enquote  [1]{``#1''}%
\providecommand \bibnamefont  [1]{#1}%
\providecommand \bibfnamefont [1]{#1}%
\providecommand \citenamefont [1]{#1}%
\providecommand \href@noop [0]{\@secondoftwo}%
\providecommand \href [0]{\begingroup \@sanitize@url \@href}%
\providecommand \@href[1]{\@@startlink{#1}\@@href}%
\providecommand \@@href[1]{\endgroup#1\@@endlink}%
\providecommand \@sanitize@url [0]{\catcode `\\12\catcode `\$12\catcode
  `\&12\catcode `\#12\catcode `\^12\catcode `\_12\catcode `\%12\relax}%
\providecommand \@@startlink[1]{}%
\providecommand \@@endlink[0]{}%
\providecommand \url  [0]{\begingroup\@sanitize@url \@url }%
\providecommand \@url [1]{\endgroup\@href {#1}{\urlprefix }}%
\providecommand \urlprefix  [0]{URL }%
\providecommand \Eprint [0]{\href }%
\providecommand \doibase [0]{https://doi.org/}%
\providecommand \selectlanguage [0]{\@gobble}%
\providecommand \bibinfo  [0]{\@secondoftwo}%
\providecommand \bibfield  [0]{\@secondoftwo}%
\providecommand \translation [1]{[#1]}%
\providecommand \BibitemOpen [0]{}%
\providecommand \bibitemStop [0]{}%
\providecommand \bibitemNoStop [0]{.\EOS\space}%
\providecommand \EOS [0]{\spacefactor3000\relax}%
\providecommand \BibitemShut  [1]{\csname bibitem#1\endcsname}%
\let\auto@bib@innerbib\@empty
%</preamble>
\bibitem [{\citenamefont {Dixon}\ \emph {et~al.}(1985)\citenamefont {Dixon},
  \citenamefont {Harvey}, \citenamefont {Vafa},\ and\ \citenamefont
  {Witten}}]{Dixon:1985jw}%
  \BibitemOpen
  \bibfield  {author} {\bibinfo {author} {\bibfnamefont {L.~J.}\ \bibnamefont
  {Dixon}}, \bibinfo {author} {\bibfnamefont {J.~A.}\ \bibnamefont {Harvey}},
  \bibinfo {author} {\bibfnamefont {C.}~\bibnamefont {Vafa}},\ and\ \bibinfo
  {author} {\bibfnamefont {E.}~\bibnamefont {Witten}},\ }\bibfield  {title}
  {\bibinfo {title} {Strings on orbifolds},\ }\href
  {https://doi.org/10.1016/0550-3213(85)90593-0} {\bibfield  {journal}
  {\bibinfo  {journal} {Nucl. Phys.}\ }\textbf {\bibinfo {volume} {B261}},\
  \bibinfo {pages} {678} (\bibinfo {year} {1985})}\BibitemShut {NoStop}%
\bibitem [{\citenamefont {Dixon}\ \emph {et~al.}(1986)\citenamefont {Dixon},
  \citenamefont {Harvey}, \citenamefont {Vafa},\ and\ \citenamefont
  {Witten}}]{Dixon:1986jc}%
  \BibitemOpen
  \bibfield  {author} {\bibinfo {author} {\bibfnamefont {L.~J.}\ \bibnamefont
  {Dixon}}, \bibinfo {author} {\bibfnamefont {J.~A.}\ \bibnamefont {Harvey}},
  \bibinfo {author} {\bibfnamefont {C.}~\bibnamefont {Vafa}},\ and\ \bibinfo
  {author} {\bibfnamefont {E.}~\bibnamefont {Witten}},\ }\bibfield  {title}
  {\bibinfo {title} {Strings on orbifolds. 2},\ }\href
  {https://doi.org/10.1016/0550-3213(86)90287-7} {\bibfield  {journal}
  {\bibinfo  {journal} {Nucl. Phys.}\ }\textbf {\bibinfo {volume} {B274}},\
  \bibinfo {pages} {285} (\bibinfo {year} {1986})}\BibitemShut {NoStop}%
\bibitem [{\citenamefont {Faraggi}\ \emph {et~al.}(1990)\citenamefont
  {Faraggi}, \citenamefont {Nanopoulos},\ and\ \citenamefont
  {Yuan}}]{Faraggi:1989ka}%
  \BibitemOpen
  \bibfield  {author} {\bibinfo {author} {\bibfnamefont {A.~E.}\ \bibnamefont
  {Faraggi}}, \bibinfo {author} {\bibfnamefont {D.~V.}\ \bibnamefont
  {Nanopoulos}},\ and\ \bibinfo {author} {\bibfnamefont {K.-j.}\ \bibnamefont
  {Yuan}},\ }\bibfield  {title} {\bibinfo {title} {{A Standard like model in
  the 4D free fermionic string formulation}},\ }\href
  {https://doi.org/10.1016/0550-3213(90)90498-3} {\bibfield  {journal}
  {\bibinfo  {journal} {Nucl. Phys.}\ }\textbf {\bibinfo {volume} {B335}},\
  \bibinfo {pages} {347} (\bibinfo {year} {1990})}\BibitemShut {NoStop}%
\bibitem [{\citenamefont {Lebedev}\ \emph {et~al.}(2007)\citenamefont {Lebedev}
  \emph {et~al.}}]{Lebedev:2006kn}%
  \BibitemOpen
  \bibfield  {author} {\bibinfo {author} {\bibfnamefont {O.}~\bibnamefont
  {Lebedev}} \emph {et~al.},\ }\bibfield  {title} {\bibinfo {title} {{A
  mini-landscape of exact MSSM spectra in heterotic orbifolds}},\ }\href
  {https://doi.org/10.1016/j.physletb.2006.12.012} {\bibfield  {journal}
  {\bibinfo  {journal} {Phys. Lett.}\ }\textbf {\bibinfo {volume} {B645}},\
  \bibinfo {pages} {88} (\bibinfo {year} {2007})},\ \Eprint
  {https://arxiv.org/abs/hep-th/0611095} {arXiv:hep-th/0611095} \BibitemShut
  {NoStop}%
\bibitem [{\citenamefont {Blaszczyk}\ \emph {et~al.}(2010)\citenamefont
  {Blaszczyk} \emph {et~al.}}]{Blaszczyk:2009in}%
  \BibitemOpen
  \bibfield  {author} {\bibinfo {author} {\bibfnamefont {M.}~\bibnamefont
  {Blaszczyk}} \emph {et~al.},\ }\bibfield  {title} {\bibinfo {title} {{A Z2xZ2
  standard model}},\ }\href {https://doi.org/10.1016/j.physletb.2009.12.036}
  {\bibfield  {journal} {\bibinfo  {journal} {Phys. Lett.}\ }\textbf {\bibinfo
  {volume} {B683}},\ \bibinfo {pages} {340} (\bibinfo {year} {2010})},\ \Eprint
  {https://arxiv.org/abs/0911.4905} {arXiv:0911.4905 [hep-th]} \BibitemShut
  {NoStop}%
\bibitem [{\citenamefont {Duff}(1990)}]{Duff:1989tf}%
  \BibitemOpen
  \bibfield  {author} {\bibinfo {author} {\bibfnamefont {M.~J.}\ \bibnamefont
  {Duff}},\ }\bibfield  {title} {\bibinfo {title} {{Duality Rotations in String
  Theory}},\ }\href {https://doi.org/10.1016/0550-3213(90)90520-N} {\bibfield
  {journal} {\bibinfo  {journal} {Nucl. Phys. B}\ }\textbf {\bibinfo {volume}
  {335}},\ \bibinfo {pages} {610} (\bibinfo {year} {1990})}\BibitemShut
  {NoStop}%
\bibitem [{\citenamefont {Strominger}\ \emph {et~al.}(1996)\citenamefont
  {Strominger}, \citenamefont {Yau},\ and\ \citenamefont
  {Zaslow}}]{Strominger:1996it}%
  \BibitemOpen
  \bibfield  {author} {\bibinfo {author} {\bibfnamefont {A.}~\bibnamefont
  {Strominger}}, \bibinfo {author} {\bibfnamefont {S.-T.}\ \bibnamefont
  {Yau}},\ and\ \bibinfo {author} {\bibfnamefont {E.}~\bibnamefont {Zaslow}},\
  }\bibfield  {title} {\bibinfo {title} {{Mirror symmetry is T duality}},\
  }\href {https://doi.org/10.1016/0550-3213(96)00434-8} {\bibfield  {journal}
  {\bibinfo  {journal} {Nucl. Phys. B}\ }\textbf {\bibinfo {volume} {479}},\
  \bibinfo {pages} {243} (\bibinfo {year} {1996})},\ \Eprint
  {https://arxiv.org/abs/hep-th/9606040} {arXiv:hep-th/9606040} \BibitemShut
  {NoStop}%
\bibitem [{\citenamefont {Dabholkar}\ and\ \citenamefont
  {Hull}(2006)}]{Dabholkar:2005ve}%
  \BibitemOpen
  \bibfield  {author} {\bibinfo {author} {\bibfnamefont {A.}~\bibnamefont
  {Dabholkar}}\ and\ \bibinfo {author} {\bibfnamefont {C.}~\bibnamefont
  {Hull}},\ }\bibfield  {title} {\bibinfo {title} {{Generalised T-duality and
  non-geometric backgrounds}},\ }\href
  {https://doi.org/10.1088/1126-6708/2006/05/009} {\bibfield  {journal}
  {\bibinfo  {journal} {JHEP}\ }\textbf {\bibinfo {volume} {05}},\ \bibinfo
  {pages} {009}},\ \Eprint {https://arxiv.org/abs/hep-th/0512005}
  {arXiv:hep-th/0512005} \BibitemShut {NoStop}%
\bibitem [{\citenamefont {Grana}\ \emph {et~al.}(2009)\citenamefont {Grana},
  \citenamefont {Minasian}, \citenamefont {Petrini},\ and\ \citenamefont
  {Waldram}}]{Grana:2008yw}%
  \BibitemOpen
  \bibfield  {author} {\bibinfo {author} {\bibfnamefont {M.}~\bibnamefont
  {Grana}}, \bibinfo {author} {\bibfnamefont {R.}~\bibnamefont {Minasian}},
  \bibinfo {author} {\bibfnamefont {M.}~\bibnamefont {Petrini}},\ and\ \bibinfo
  {author} {\bibfnamefont {D.}~\bibnamefont {Waldram}},\ }\bibfield  {title}
  {\bibinfo {title} {{T-duality, Generalized Geometry and Non-Geometric
  Backgrounds}},\ }\href {https://doi.org/10.1088/1126-6708/2009/04/075}
  {\bibfield  {journal} {\bibinfo  {journal} {JHEP}\ }\textbf {\bibinfo
  {volume} {04}},\ \bibinfo {pages} {075}},\ \Eprint
  {https://arxiv.org/abs/0807.4527} {arXiv:0807.4527 [hep-th]} \BibitemShut
  {NoStop}%
\bibitem [{\citenamefont {Narain}\ \emph {et~al.}(1987)\citenamefont {Narain},
  \citenamefont {Sarmadi},\ and\ \citenamefont {Vafa}}]{narain_87}%
  \BibitemOpen
  \bibfield  {author} {\bibinfo {author} {\bibfnamefont {K.~S.}\ \bibnamefont
  {Narain}}, \bibinfo {author} {\bibfnamefont {M.~H.}\ \bibnamefont
  {Sarmadi}},\ and\ \bibinfo {author} {\bibfnamefont {C.}~\bibnamefont
  {Vafa}},\ }\bibfield  {title} {\bibinfo {title} {Asymmetric orbifolds},\
  }\href@noop {} {\bibfield  {journal} {\bibinfo  {journal} {Nucl. Phys.}\
  }\textbf {\bibinfo {volume} {B288}},\ \bibinfo {pages} {551} (\bibinfo {year}
  {1987})}\BibitemShut {NoStop}%
\bibitem [{\citenamefont {Ib{\'a}{\~n}ez}\ \emph
  {et~al.}(1988{\natexlab{a}})\citenamefont {Ib{\'a}{\~n}ez}, \citenamefont
  {Mas}, \citenamefont {Nilles},\ and\ \citenamefont {Quevedo}}]{ibanez_88}%
  \BibitemOpen
  \bibfield  {author} {\bibinfo {author} {\bibfnamefont {L.~E.}\ \bibnamefont
  {Ib{\'a}{\~n}ez}}, \bibinfo {author} {\bibfnamefont {J.}~\bibnamefont {Mas}},
  \bibinfo {author} {\bibfnamefont {H.~P.}\ \bibnamefont {Nilles}},\ and\
  \bibinfo {author} {\bibfnamefont {F.}~\bibnamefont {Quevedo}},\ }\bibfield
  {title} {\bibinfo {title} {Heterotic strings in symmetric and asymmetric
  orbifold backgrounds},\ }\href@noop {} {\bibfield  {journal} {\bibinfo
  {journal} {Nucl. Phys.}\ }\textbf {\bibinfo {volume} {B301}},\ \bibinfo
  {pages} {157} (\bibinfo {year} {1988}{\natexlab{a}})}\BibitemShut {NoStop}%
\bibitem [{\citenamefont {Di~Francesco}\ \emph {et~al.}(1997)\citenamefont
  {Di~Francesco}, \citenamefont {Mathieu},\ and\ \citenamefont
  {Senechal}}]{DiFrancesco:1997nk}%
  \BibitemOpen
  \bibfield  {author} {\bibinfo {author} {\bibfnamefont {P.}~\bibnamefont
  {Di~Francesco}}, \bibinfo {author} {\bibfnamefont {P.}~\bibnamefont
  {Mathieu}},\ and\ \bibinfo {author} {\bibfnamefont {D.}~\bibnamefont
  {Senechal}},\ }\href@noop {} {\emph {\bibinfo {title} {Conformal field
  theory}}}\ (\bibinfo  {publisher} {Springer},\ \bibinfo {address} {New York,
  USA},\ \bibinfo {year} {1997})\ \bibinfo {note} {890 p}\BibitemShut {NoStop}%
\bibitem [{\citenamefont {Athanasopoulos}\ \emph {et~al.}(2016)\citenamefont
  {Athanasopoulos}, \citenamefont {Faraggi}, \citenamefont {Groot~Nibbelink},\
  and\ \citenamefont {Mehta}}]{Athanasopoulos:2016aws}%
  \BibitemOpen
  \bibfield  {author} {\bibinfo {author} {\bibfnamefont {P.}~\bibnamefont
  {Athanasopoulos}}, \bibinfo {author} {\bibfnamefont {A.~E.}\ \bibnamefont
  {Faraggi}}, \bibinfo {author} {\bibfnamefont {S.}~\bibnamefont
  {Groot~Nibbelink}},\ and\ \bibinfo {author} {\bibfnamefont {V.~M.}\
  \bibnamefont {Mehta}},\ }\bibfield  {title} {\bibinfo {title} {{Heterotic
  free fermionic and symmetric toroidal orbifold models}},\ }\href
  {https://doi.org/10.1007/JHEP04(2016)038} {\bibfield  {journal} {\bibinfo
  {journal} {JHEP}\ }\textbf {\bibinfo {volume} {04}},\ \bibinfo {pages}
  {038}},\ \Eprint {https://arxiv.org/abs/1602.03082} {arXiv:1602.03082
  [hep-th]} \BibitemShut {NoStop}%
\bibitem [{\citenamefont {Faraggi}(1994)}]{Faraggi:1994cv}%
  \BibitemOpen
  \bibfield  {author} {\bibinfo {author} {\bibfnamefont {A.~E.}\ \bibnamefont
  {Faraggi}},\ }\bibfield  {title} {\bibinfo {title} {{Proton stability in
  superstring derived models}},\ }\href
  {https://doi.org/10.1016/0550-3213(94)90194-5} {\bibfield  {journal}
  {\bibinfo  {journal} {Nucl. Phys. B}\ }\textbf {\bibinfo {volume} {428}},\
  \bibinfo {pages} {111} (\bibinfo {year} {1994})},\ \Eprint
  {https://arxiv.org/abs/hep-ph/9403312} {arXiv:hep-ph/9403312} \BibitemShut
  {NoStop}%
\bibitem [{\citenamefont {Faraggi}(2001)}]{dtsm}%
  \BibitemOpen
  \bibfield  {author} {\bibinfo {author} {\bibfnamefont {A.~E.}\ \bibnamefont
  {Faraggi}},\ }\bibfield  {title} {\bibinfo {title} {Doublet--triplet
  splitting in realistic heterotic string derived models},\ }\href
  {https://doi.org/https://doi.org/10.1016/S0370-2693(01)01165-0} {\bibfield
  {journal} {\bibinfo  {journal} {Physics Letters B}\ }\textbf {\bibinfo
  {volume} {520}},\ \bibinfo {pages} {337} (\bibinfo {year}
  {2001})}\BibitemShut {NoStop}%
\bibitem [{\citenamefont {Faraggi}(1993)}]{yukawa}%
  \BibitemOpen
  \bibfield  {author} {\bibinfo {author} {\bibfnamefont {A.~E.}\ \bibnamefont
  {Faraggi}},\ }\bibfield  {title} {\bibinfo {title} {Yukawa couplings in
  superstring-derived standard-like models},\ }\href
  {https://link.aps.org/doi/10.1103/PhysRevD.47.5021} {\bibfield  {journal}
  {\bibinfo  {journal} {Phys. Rev. D}\ }\textbf {\bibinfo {volume} {47}},\
  \bibinfo {pages} {5021} (\bibinfo {year} {1993})}\BibitemShut {NoStop}%
\bibitem [{\citenamefont {Faraggi}(2005)}]{moduli}%
  \BibitemOpen
  \bibfield  {author} {\bibinfo {author} {\bibfnamefont {A.~E.}\ \bibnamefont
  {Faraggi}},\ }\bibfield  {title} {\bibinfo {title} {Moduli fixing in
  realistic string vacua},\ }\href
  {https://doi.org/https://doi.org/10.1016/j.nuclphysb.2005.08.028} {\bibfield
  {journal} {\bibinfo  {journal} {Nuclear Physics B}\ }\textbf {\bibinfo
  {volume} {728}},\ \bibinfo {pages} {83} (\bibinfo {year} {2005})}\BibitemShut
  {NoStop}%
\bibitem [{\citenamefont {Gross}\ \emph {et~al.}(1985)\citenamefont {Gross},
  \citenamefont {Harvey}, \citenamefont {Martinec},\ and\ \citenamefont
  {Rohm}}]{Gross:1985fr}%
  \BibitemOpen
  \bibfield  {author} {\bibinfo {author} {\bibfnamefont {D.~J.}\ \bibnamefont
  {Gross}}, \bibinfo {author} {\bibfnamefont {J.~A.}\ \bibnamefont {Harvey}},
  \bibinfo {author} {\bibfnamefont {E.~J.}\ \bibnamefont {Martinec}},\ and\
  \bibinfo {author} {\bibfnamefont {R.}~\bibnamefont {Rohm}},\ }\bibfield
  {title} {\bibinfo {title} {Heterotic string theory. 1. {T}he free heterotic
  string},\ }\href {https://doi.org/10.1016/0550-3213(85)90394-3} {\bibfield
  {journal} {\bibinfo  {journal} {Nucl. Phys.}\ }\textbf {\bibinfo {volume}
  {B256}},\ \bibinfo {pages} {253} (\bibinfo {year} {1985})}\BibitemShut
  {NoStop}%
\bibitem [{\citenamefont {Ib{\'a}{\~n}ez}\ \emph
  {et~al.}(1988{\natexlab{b}})\citenamefont {Ib{\'a}{\~n}ez}, \citenamefont
  {Mas}, \citenamefont {Nilles},\ and\ \citenamefont
  {Quevedo}}]{Ibanez:1987pj}%
  \BibitemOpen
  \bibfield  {author} {\bibinfo {author} {\bibfnamefont {L.~E.}\ \bibnamefont
  {Ib{\'a}{\~n}ez}}, \bibinfo {author} {\bibfnamefont {J.}~\bibnamefont {Mas}},
  \bibinfo {author} {\bibfnamefont {H.-P.}\ \bibnamefont {Nilles}},\ and\
  \bibinfo {author} {\bibfnamefont {F.}~\bibnamefont {Quevedo}},\ }\bibfield
  {title} {\bibinfo {title} {Heterotic strings in symmetric and asymmetric
  orbifold backgrounds},\ }\href@noop {} {\bibfield  {journal} {\bibinfo
  {journal} {Nucl. Phys.}\ }\textbf {\bibinfo {volume} {B301}},\ \bibinfo
  {pages} {157} (\bibinfo {year} {1988}{\natexlab{b}})}\BibitemShut {NoStop}%
\bibitem [{\citenamefont {Groot~Nibbelink}\ and\ \citenamefont
  {Vaudrevange}(2017)}]{GrootNibbelink:2017usl}%
  \BibitemOpen
  \bibfield  {author} {\bibinfo {author} {\bibfnamefont {S.}~\bibnamefont
  {Groot~Nibbelink}}\ and\ \bibinfo {author} {\bibfnamefont {P.~K.~S.}\
  \bibnamefont {Vaudrevange}},\ }\bibfield  {title} {\bibinfo {title}
  {{T-duality orbifolds of heterotic Narain compactifications}},\ }\href
  {https://doi.org/10.1007/JHEP04(2017)030} {\bibfield  {journal} {\bibinfo
  {journal} {JHEP}\ }\textbf {\bibinfo {volume} {04}},\ \bibinfo {pages}
  {030}},\ \Eprint {https://arxiv.org/abs/1703.05323} {arXiv:1703.05323
  [hep-th]} \BibitemShut {NoStop}%
\bibitem [{\citenamefont {Groot~Nibbelink}(2021)}]{GrootNibbelink:2020dib}%
  \BibitemOpen
  \bibfield  {author} {\bibinfo {author} {\bibfnamefont {S.}~\bibnamefont
  {Groot~Nibbelink}},\ }\bibfield  {title} {\bibinfo {title} {{A worldsheet
  perspective on heterotic T-duality orbifolds}},\ }\href
  {https://doi.org/10.1007/JHEP04(2021)190} {\bibfield  {journal} {\bibinfo
  {journal} {JHEP}\ }\textbf {\bibinfo {volume} {04}},\ \bibinfo {pages}
  {190}},\ \Eprint {https://arxiv.org/abs/2012.02778} {arXiv:2012.02778
  [hep-th]} \BibitemShut {NoStop}%
\bibitem [{\citenamefont {Antoniadis}\ \emph {et~al.}(1987)\citenamefont
  {Antoniadis}, \citenamefont {Bachas},\ and\ \citenamefont
  {Kounnas}}]{Antoniadis1987a}%
  \BibitemOpen
  \bibfield  {author} {\bibinfo {author} {\bibfnamefont {I.}~\bibnamefont
  {Antoniadis}}, \bibinfo {author} {\bibfnamefont {C.}~\bibnamefont {Bachas}},\
  and\ \bibinfo {author} {\bibfnamefont {C.}~\bibnamefont {Kounnas}},\
  }\bibfield  {title} {\bibinfo {title} {{Four-dimensional superstrings}},\
  }\href {https://doi.org/10.1016/0550-3213(87)90372-5} {\bibfield  {journal}
  {\bibinfo  {journal} {Nucl. Phys.}\ }\textbf {\bibinfo {volume} {B289}},\
  \bibinfo {pages} {87} (\bibinfo {year} {1987})}\BibitemShut {NoStop}%
\bibitem [{\citenamefont {Kawai}\ \emph {et~al.}(1987)\citenamefont {Kawai},
  \citenamefont {Lewellen},\ and\ \citenamefont {{Henry Tye}}}]{Kawai1987}%
  \BibitemOpen
  \bibfield  {author} {\bibinfo {author} {\bibfnamefont {H.}~\bibnamefont
  {Kawai}}, \bibinfo {author} {\bibfnamefont {D.~C.}\ \bibnamefont
  {Lewellen}},\ and\ \bibinfo {author} {\bibfnamefont {S.-H.}\ \bibnamefont
  {{Henry Tye}}},\ }\bibfield  {title} {\bibinfo {title} {{Construction of
  fermionic string models in four dimensions}},\ }\href
  {https://doi.org/10.1016/0550-3213(87)90208-2} {\bibfield  {journal}
  {\bibinfo  {journal} {Nucl. Phys.}\ }\textbf {\bibinfo {volume} {B288}},\
  \bibinfo {pages} {1} (\bibinfo {year} {1987})}\BibitemShut {NoStop}%
\bibitem [{\citenamefont {Antoniadis}\ and\ \citenamefont {Bachas}(1988)}]{AB}%
  \BibitemOpen
  \bibfield  {author} {\bibinfo {author} {\bibfnamefont {I.}~\bibnamefont
  {Antoniadis}}\ and\ \bibinfo {author} {\bibfnamefont {C.}~\bibnamefont
  {Bachas}},\ }\bibfield  {title} {\bibinfo {title} {4d fermionic superstrings
  with arbitrary twists},\ }\href
  {https://doi.org/10.1016/0550-3213(88)90355-0} {\bibfield  {journal}
  {\bibinfo  {journal} {Nucl. Phys.}\ }\textbf {\bibinfo {volume} {B298}},\
  \bibinfo {pages} {586} (\bibinfo {year} {1988})}\BibitemShut {NoStop}%
\bibitem [{\citenamefont {Ferrara}\ \emph {et~al.}(1987)\citenamefont
  {Ferrara}, \citenamefont {Girardello}, \citenamefont {Kounnas},\ and\
  \citenamefont {Porrati}}]{Ferrara:1987jr}%
  \BibitemOpen
  \bibfield  {author} {\bibinfo {author} {\bibfnamefont {S.}~\bibnamefont
  {Ferrara}}, \bibinfo {author} {\bibfnamefont {L.}~\bibnamefont {Girardello}},
  \bibinfo {author} {\bibfnamefont {C.}~\bibnamefont {Kounnas}},\ and\ \bibinfo
  {author} {\bibfnamefont {M.}~\bibnamefont {Porrati}},\ }\bibfield  {title}
  {\bibinfo {title} {{The Effective Interactions of Chiral Families in
  Four-dimensional Superstrings}},\ }\href
  {https://doi.org/10.1016/0370-2693(87)91066-5} {\bibfield  {journal}
  {\bibinfo  {journal} {Phys. Lett. B}\ }\textbf {\bibinfo {volume} {194}},\
  \bibinfo {pages} {358} (\bibinfo {year} {1987})}\BibitemShut {NoStop}%
\bibitem [{\citenamefont {Antoniadis}\ \emph {et~al.}(1988)\citenamefont
  {Antoniadis}, \citenamefont {Ellis}, \citenamefont {Hagelin},\ and\
  \citenamefont {Nanopoulos}}]{Antoniadis:1987tv}%
  \BibitemOpen
  \bibfield  {author} {\bibinfo {author} {\bibfnamefont {I.}~\bibnamefont
  {Antoniadis}}, \bibinfo {author} {\bibfnamefont {J.~R.}\ \bibnamefont
  {Ellis}}, \bibinfo {author} {\bibfnamefont {J.~S.}\ \bibnamefont {Hagelin}},\
  and\ \bibinfo {author} {\bibfnamefont {D.~V.}\ \bibnamefont {Nanopoulos}},\
  }\bibfield  {title} {\bibinfo {title} {{GUT Model Building with Fermionic
  Four-Dimensional Strings}},\ }\href
  {https://doi.org/10.1016/0370-2693(88)90978-1} {\bibfield  {journal}
  {\bibinfo  {journal} {Phys. Lett. B}\ }\textbf {\bibinfo {volume} {205}},\
  \bibinfo {pages} {459} (\bibinfo {year} {1988})}\BibitemShut {NoStop}%
\bibitem [{\citenamefont {Faraggi}\ and\ \citenamefont
  {Nanopoulos}(1993)}]{Faraggi1993}%
  \BibitemOpen
  \bibfield  {author} {\bibinfo {author} {\bibfnamefont {A.~E.}\ \bibnamefont
  {Faraggi}}\ and\ \bibinfo {author} {\bibfnamefont {D.~V.}\ \bibnamefont
  {Nanopoulos}},\ }\bibfield  {title} {\bibinfo {title} {{Naturalness of three
  generations in free fermionic $Z_2^n\otimes Z_4$ string models}},\ }\href
  {https://doi.org/10.1103/PhysRevD.48.3288} {\bibfield  {journal} {\bibinfo
  {journal} {Phys. Rev.}\ }\textbf {\bibinfo {volume} {D48}},\ \bibinfo {pages}
  {3288} (\bibinfo {year} {1993})}\BibitemShut {NoStop}%
\bibitem [{\citenamefont {Faraggi}(1992)}]{slm1}%
  \BibitemOpen
  \bibfield  {author} {\bibinfo {author} {\bibfnamefont {A.~E.}\ \bibnamefont
  {Faraggi}},\ }\bibfield  {title} {\bibinfo {title} {A new standard-like model
  in the four dimensional free fermionic string formulation},\ }\href
  {https://doi.org/https://doi.org/10.1016/0370-2693(92)90723-H} {\bibfield
  {journal} {\bibinfo  {journal} {Physics Letters B}\ }\textbf {\bibinfo
  {volume} {278}},\ \bibinfo {pages} {131} (\bibinfo {year}
  {1992})}\BibitemShut {NoStop}%
\bibitem [{\citenamefont {Cleaver}\ \emph {et~al.}(2008)\citenamefont
  {Cleaver}, \citenamefont {Faraggi}, \citenamefont {Manno},\ and\
  \citenamefont {Timirgaziu}}]{cleaverwithmanno}%
  \BibitemOpen
  \bibfield  {author} {\bibinfo {author} {\bibfnamefont {G.~B.}\ \bibnamefont
  {Cleaver}}, \bibinfo {author} {\bibfnamefont {A.~E.}\ \bibnamefont
  {Faraggi}}, \bibinfo {author} {\bibfnamefont {E.}~\bibnamefont {Manno}},\
  and\ \bibinfo {author} {\bibfnamefont {C.}~\bibnamefont {Timirgaziu}},\
  }\bibfield  {title} {\bibinfo {title} {Quasi-realistic heterotic-string
  models with vanishing one-loop cosmological constant and perturbatively
  broken supersymmetry?},\ }\href {https://doi.org/10.1103/PhysRevD.78.046009}
  {\bibfield  {journal} {\bibinfo  {journal} {Phys. Rev. D}\ }\textbf {\bibinfo
  {volume} {78}},\ \bibinfo {pages} {046009} (\bibinfo {year}
  {2008})}\BibitemShut {NoStop}%
\bibitem [{\citenamefont {Faraggi}\ \emph {et~al.}(2020)\citenamefont
  {Faraggi}, \citenamefont {Matyas},\ and\ \citenamefont {Percival}}]{stable}%
  \BibitemOpen
  \bibfield  {author} {\bibinfo {author} {\bibfnamefont {A.}~\bibnamefont
  {Faraggi}}, \bibinfo {author} {\bibfnamefont {V.}~\bibnamefont {Matyas}},\
  and\ \bibinfo {author} {\bibfnamefont {B.}~\bibnamefont {Percival}},\
  }\bibfield  {title} {\bibinfo {title} {Stable three generation standard-like
  model from a tachyonic ten dimensional heterotic-string vacuum},\ }\bibfield
  {journal} {\bibinfo  {journal} {The European Physical Journal C}\ }\textbf
  {\bibinfo {volume} {80}},\ \href
  {https://doi.org/10.1140/epjc/s10052-020-7894-x}
  {10.1140/epjc/s10052-020-7894-x} (\bibinfo {year} {2020})\BibitemShut
  {NoStop}%
\end{thebibliography}%

\end{document}